\documentstyle[multicol,aps,epsf]{revtex}

\begin{document}

\title{Conductivity of the classical two-dimensional electron gas}

\author{M.Hruska, B.Spivak}

\address{Physics Department, University of Washington, Seattle, WA 98195}

\maketitle 

\begin{abstract} 

We discuss the applicability of the Boltzmann equation to the
classical two-dimensional electron gas.
We show that in the presence of both the electron-impurity and
electron-electron
scattering the Boltzmann equation can be inapplicable and the correct
result for conductivity can be different from the one obtained from 
the kinetic equation by a logarithmically
large factor.
\end{abstract}


\begin{multicols}{2}

The description of kinetics of the classical electron gas in
 semiconductors is usually done with the help of the Boltzmann kinetic
equation \cite{sarma}
\begin{eqnarray} 
\frac{\partial n_{\bbox{p}}(\bbox{r},t)}{\partial t}+\bbox{v} \cdot
\frac{\partial n_{\bbox p}(\bbox r,t)}{\partial
\bbox{r}}+ e \bbox E
\cdot \frac{\partial n_{\bbox p}(\bbox r,t)}{\partial \bbox p}  && \nonumber\\
= I_{imp}(n_{\bbox p}(\bbox r,t))+I_{ee}(n_{\bbox p}(\bbox r,t))+I_{ph}(n_{\bbox p}(\bbox r,t)) && \label{eq1}
\end{eqnarray}
where $n_{\bbox{p}}(\bbox{r},t)$ is the electron distribution function,
$\bbox{E}$ is the
external electric field, $\bbox v$ and
$\bbox{p}$ are the electron velocity and momentum, and $I_{imp},I_{ee}$
and $I_{ph}$ are the electron-impurity,
 electron-electron and electron-phonon scattering integrals 
which are characterized by the relaxation times $\tau_{i}$, $\tau_{e}$ and
 $\tau_{ph}$ respectively.

At low enough temperatures $T$ when $\tau_{i}\ll \tau_{ph},\tau_{e}$,
the conductivity of the sample
is determined by the electron-impurity scatterings and
is given by the Drude formula
\begin{equation}\label{eq2}
 \sigma_{0}=e^{2}D\nu 
\end{equation}
Here  $D=v_F^2 \tau_{i} /2$ (where $v_F$ is the Fermi velocity, $\tau _i
=1/v_F N_i a$, $N_i$ is the impurity
concentration  and $a$ is the amplitude of the electron-impurity
scattering), and $\nu$ is the electron density of states
at the Fermi level. To be concrete, we consider the case of a degenerate
electron gas, though our results are of a general character.
Eq.~(\ref{eq2}) also holds in the case of intermediate temperatures
when $\tau_{ph}\gg 
\tau_{i}\gg \tau_{e}$. Indeed, since the electron-electron
scatterings 
conserve the total momentum in the electron system, they do not
contribute directly to the resistance of the system. 

 The form of $I_{imp}$ in Eq.~(\ref{eq1}) is a result of a procedure in
which
 averaging over realizations of the random impurity potential is done
before solving the equation of motion.
The criteria of applicability of the conventional Boltzmann
 kinetic equation~(\ref{eq1}) correspond to the absence of correlations
between subsequent scattering events. This leads us to a requirement that
the electron
 gas is weakly interacting 
\begin{equation}\label{eq3}
 \tau_{i},\tau_{e}\gg t_{coll}
 \end{equation}
 where $t_{coll}$ is the the collision time, which is estimated as $a/v$ 
in classical mechanics,
and as $\frac{\hbar}{\epsilon}$ in quantum mechanics  ($\epsilon$
is the characteristic electron energy).
 
It has been shown \cite{abr} that in the two-dimensional case even under
 conditions~(\ref{eq3})  the quantum interference corrections to the
Drude
 formula logarithmically diverge  as temperature $T$ goes to
zero. 
  This means that in this case the kinetic equation~(\ref{eq1}) is not
applicable at small $T$.
  
  In this article we study the case of high enough temperatures when the
 quantum interference corrections can be neglected. The classical
corrections to Eq.~(\ref{eq2}) associated with correlations between
subsequent scatterings
 were considered in
\cite{vanLeeuwen,peierls} in the framework of the Lorentz gas model. It
was shown that in the two-dimensional case
correlational corrections to the diffusion
coefficient and to the conductivity are small, but nonanalytical functions
of the concentration of scatterers $N_{i}$. 

We consider in this article the case of a two-dimensional electron gas
with the mean free path $l_e =v_F \tau _e$ much smaller than the average
distance between the impurities: $l_e \ll N_i^{-1/2}$. We show that under
conditions~(\ref{eq3}) a
combination of electron-electron and electron-impurity scatterings can 
lead to big correlational corrections to the Drude formula and that
  the correct expression for the conductivity of the system is
  \begin{equation}\label{eq4}
  \sigma = \sigma_0 \big(1+ \frac{a v_F}{2 \pi \eta } \ln
\frac{1}{N^{1/2}_{i} l_e}\big)
  \end{equation}
where $\eta$ is the kinematic viscosity of the electron liquid, which is
estimated for a Fermi liquid as $\eta \sim v_F l_e$. 
  Eq.~(\ref{eq4}) differs from Eq.~(\ref{eq2}) by a logarithmically large
factor, provided
\[
\frac{a v_F}{2 \pi \eta} \ln \frac{1}{N^{1/2}_{i} l_e}\gg 1.
\]
  
  On distances from impurities larger than $\l_{e}$, the motion of the
 electron system can be described by the Navier-Stokes hydrodynamic
equation 
  which can be derived from Eq.~(\ref{eq1})  with
$I_{imp}=0$ \cite{pitaevski}.
In this case the elastic scatterers should be described by a random
potential and the averaging over the random realizations of the potential
should be done after solving the Navier-Stokes
equation. This should in principle give a result for conductivity that is 
different from the one implied by Eq.~(\ref{eq1}).
The two approaches are equivalent  in the
approximation which assumes the
variation of the hydrodynamic velocity between impurities to be small.

The idea of this paper is that, in fact, in the two-dimensional case the
hydrodynamic
velocity of 
electrons is a significantly nonuniform function of coordinates: 
 the velocity near impurities
 is much smaller  than inbetween them.
  This is in contrast to the three-dimensional case where the spatial
variation of
 the velocity is small and Eqs.~(\ref{eq1}),~(\ref{eq2}) are
 applicable. 
  The reason for the significant nonuniformity of the velocity field is
connected to the Stokes paradox \cite{landau}:  in the
presence
of an obstacle bypassed by the fluid, the 
solution of the linearized Navier-Stokes equation in the two-dimensional 
case grows logarithmically
with the
distance from the obstacle. This means the absence of a bounded solution.  
In our case  the finite concentration of scatterers provides a bound on
the
magnitude of the velocity field. However, the
magnitude 
of the average velocity of electrons induced by the electric field turns
out to be logarithmically bigger than that which follows from
Eq.~(\ref{eq1}).

In the approximation linear in $\bbox E$, the Navier-Stokes equation
has the form 
\begin{equation}\label{eq5}
-\nabla p +
 \mu \Delta \bbox u +n e \bbox E -\sum_i \bbox{f} \delta(\bbox r- \bbox r_i) =0
\end{equation}
Here $n$ is the electron concentration, $\mu = m n \eta$ is the
dynamic viscosity of the electron fluid (where $m$ is the electron
mass), and $\bbox r_i$ are coordinates of impurities labeled by the index
"i".
 The last term in Eq.~(\ref{eq5}) is associated with 
the force on the fluid by the
 i-th impurity, whose radius is much
smaller than $l_{e}$. 
Therefore the value of the force $\bbox f$ can be estimated 
 assuming that to zeroth order in $a/l_{e}$ the
distribution
function of the electrons near the $i$-th impurity ($|r-r_{i}|\leq l_{e}$)
is an
equilibrium one with the average velocity $\bbox{u}(\bbox r_{i})$.
 (In the following we will assume that all $\bbox{u}(\bbox{r}_{i})\sim
\bbox{u}_{0}$ are of the same order.)
 Consequently we have 
\cite{pitaevski}
\begin{equation}\label{eqFkin}
\bbox{f}=n p_F a \bbox u_0 . 
\end{equation}

In the stationary case the conservation of the total
momentum of the system gives 
\begin{equation}\label{eqEu0}
e \bbox E=\bbox f N_i / n.
\end{equation}

Far from the position of impurities 
($|\bbox{r}-\bbox{r}_{i}|\gg l_{e}$) the electron liquid can be 
considered as an incompressible one. 
The continuity equation for an incompressible liquid
\[
	\nabla \cdot \bbox u =0
\]
and the Navier-Stokes equation ~(\ref{eq5}) have a
solution for the pressure and the velocity
\begin{eqnarray}
\bbox u(\bbox{r})& =& C\hat {x}+
\hat x \frac{C_0}{2} \ln r -\frac{C_0}{4} (\hat r \cos\theta  +
 \hat {\theta }\sin\theta ) \nonumber\\
p(r)&=& p_0 -\mu C_{0}\frac{\cos\theta }{r}   \label{vp}
  \end{eqnarray}
where  $C_0 =\frac{f}{2\pi \mu}$, $\hat r$ and $\hat \theta$ are unit
basis vectors of the cylindrical coordinate system and the x-axis (with
the unit vector $\hat x$) is oriented parallel to the electric field.
 
 The value of the coefficient $C$ is determined 
using $\bbox u(r=l_e,\theta =\pi) = \bbox u_0$ and
Eqs.~(\ref{eqFkin}),~(\ref{eqEu0}) and~(\ref{vp}):
\[
  C= \frac{\sigma _0 E}{ ne} (1- \frac{a v_F}{ 4\pi \eta }\ln l_e) .
\]

As a result, the spatially averaged velocity $\bbox {\bar u}$ of the
electronic flow is 
\[
 \bbox {\bar u} = \frac{\sigma _0 \bbox E}{ne}(1+\frac{a v_F}{2 \pi \eta}
\ln \frac{1}{N_i^{1/2} l_e}).
\]
Expressing the current density as $\bbox j =ne \bar{\bbox u}$, the
electric conductivity Eq.~(\ref{eq4}) is obtained.

The calculations presented above assumed that the viscosity of the
electron
gas is a well-defined quantity. On the other hand it is well known that 
the fluctuational corrections to the viscosity of a 2D neutral liquid
diverge 
\cite{resibois}, which means that  hydrodynamic
equations are nonlocal in two dimensions. This fact was pointed out in
\cite{alder} and 
discussed in the most complete form in \cite{andreev80}.  Below we present a
trivial 
generalization of the procedure \cite{andreev80} to estimate
fluctuational
corrections
to the viscosity of the two-dimensional charged electron liquid.
In this case the low-lying collective excitations 
 are plasmons and the fluctuations of the charge and of the hydrodynamic
velocity can be described as a combination of plasmon excitations with 
different wave vectors $\bbox{q}$. 
The real part of the plasmon frequency is \cite{fetter} 
$\omega _{\bbox q}=\sqrt{2\pi n e^2 |\bbox{q}| /  m}$ , where $\bbox{q}$
is the
plasmon momentum. 
 Since the imaginary part of the frequency
$\omega' _{\bbox q} =\eta |\bbox{q}|^{2}/2$ is proportional to $|\bbox q|
^2$ ($\eta = \mu /m n$ is the kinematic viscosity of the liquid),
fluctuations with sufficiently low $|\bbox q|$ have arbitrarily large mean
free path and therefore plasmons are well-defined excitations. This is 
the physical reason why the
fluctuational mechanism of viscosity is always the basic one
 at sufficiently small gradients and in samples with no elastically
scattering potential.
Since the phases of the thermal fluctuations are random it is convenient
to
introduce a distribution function of plasmon fluctuations
$n(\bbox{q})$, normalized in such a way that the quantity
$\int \frac{d\bbox q}{(2\pi)^2} \omega(\bbox q) n(\bbox{q})$ is the
average energy density of fluctuations.
The flux of momentum associated with these excitations is \cite{gurevich}
\begin{equation}\label{eq15}
\Pi_{ij}=-\int \frac{d\bbox{q}}{(2\pi)^2}  q_j \frac{\partial
 \omega_{\bbox q}}{\partial {q_i}} n(\bbox q)
\end{equation}
The part of the distribution function of plasmons which is proportional to 
the gradient of the hydrodynamic velocity can be retrieved
from the kinetic equation \cite{andreev80}
\begin{equation}\label{eq14}
\frac{\partial \omega _{\bbox q}}{\partial \bbox{q}} \cdot \frac{\partial
n_{\bbox u}(\bbox q)}{\partial \bbox{r}}=-(n(\bbox r,\bbox
q)-n_{u}(\bbox q))/\tau
\end{equation}
where $\tau=\omega'^{-1}$ is the plasmon mean free time and 
 $n_{\bbox u}(\bbox q)$ is the equilibrium Bose distribution
function corresponding to a local hydrodynamic velocity $\bbox u(\bbox
r)$. At small $\omega$ and $|\bbox u|$ 
one can approximate $n_{\bbox u}(\bbox q) \simeq
\frac{T}{\omega _{\bbox q}-\bbox{q} \cdot \bbox{u}(r)}$.
Using Eq.~(\ref{eq14}) and expanding $n_{\bbox u}(\bbox q)$ in $\bbox u$, 
one can estimate from Eq.~(\ref{eq15}) the fluctuational contribution to
the viscosity  
\begin{equation}\label{eq16}
 \delta \eta \sim -\frac{1}{m n} 
 \int (q \frac{\partial \omega _{\bbox q}}{\partial q})^2 
\tau \frac{\partial n_0(\bbox q)}{\partial \omega_{\bbox q}} q dq
\end{equation}
(where $n_0(\bbox q) = n_{\bbox u = 0} (\bbox q))$. Thus the fluctuational
correction to the viscosity
can be described only in terms of hydrodynamic quantities: viscosity
and plasmon dispersion relation. In two-dimensional pure systems the
expression Eq.~(\ref{eq16}) logarithmically 
diverges. However, in the presence of the elastically scattering potential 
the dispersion relation of plasmons which we used holds only at
those frequencies for which the plasmon mean free time is shorter than the
electron-impurity scattering time
 $\tau < \tau _i$. 
This determines a cut-off in Eq.~(\ref{eq16}). As a result, the
fluctuational correction 
to the viscosity can be neglected provided
\[
	\delta \eta \sim \frac{T}{\mu }
\ln (\frac{T}{E_F}\frac{\sigma _0}{e^2 / \hbar} ) \ll \eta .
\]
 
In conclusion, we would like to stress that the existence of the large 
logarithmic factor in the Eq.~(\ref{eq4}) 
and the consequent inapplicability of the Boltzmann kinetic equation is a 
property of the two-dimensional problem.
In the three-dimensional case the Stokes paradox does not exist, the
difference between $\bbox{u}_0$ and $ \bbox{ \bar u}$ is small 
and the Boltzmann kinetic equation can be applied.

This work was partially supported by the Division of Material Sciences,
U.S.National Science Foundation under Contract No. DMR-9205144.

We would like to thank A. Zyuzin, A.Andreev, S.Chakrovarty,
M.Feigelman, A.I.Larkin, D.Khmelnitskii, S.Kivelson, F.Zhou and P.Kovtun
for valuable discussions.

\end{multicols}

\end{document}